\documentclass{epl}
\usepackage{amssymb}
\usepackage{graphicx}
\usepackage{epsfig}
\title{An invariance property of diffusive random walks}
\author{BLANCO Stephane \inst{1} \and FOURNIER Richard \inst{1}}
\institute{
  \inst{1} Laboratoire d'Energ\'etique  - Universit\'e Paul Sabatier, 
118 route de Narbonne, 31062 Toulouse Cedex 4
}
\begin{document}

\maketitle

\begin{abstract}
Starting from a simple animal-biology example,
a general, somewhat counter-intuitive property
of diffusion random walks is presented.
It is shown that for any (non-homogeneous) purely diffusing system,
under any isotropic uniform incidence,
the average length of trajectories through the system
(the average length of the random walk trajectories
from entry point to first exit point)
is independent of the characteristics of the diffusion process
and therefore depends only on the geometry of the system.
This exact invariance property may be seen as a generalization to diffusion of
the well known mean-chord-length property \cite{Case67}, leading
to broad physics and biology applications.
\end{abstract}

%\section{Section title}

Let us first consider a practical animal-biology example that was at the
origin of the theoretical work reported hereafter. It is well
established, for diverse species, that the spontaneous displacement of
insects such as ants, on an horizontal
planar surface, may be accurately modeled as a constant-speed diffusion
random-walk\cite{Turchin91,Crist91,Holmes93}.
We assume that a circle of radius $R$ is drawn on the surface and we
measure the average time that ants spend inside the circle when they
enter it (see Fig. 1). The last assumption is that measurements are
performed a long time after ants were dropped on the surface, so that
the memory of their initial position is lost~: this is enough to
ensure that no specific direction is favored and therefore
that when ants encounter the circle, their incident directions are
distributed isotropically.
Simple Monte Carlo simulations of such experiments
(see Fig. 2) indicate without any doubt
that the average encounter time (time between entry into the circle and
first exit), for a fixed velocity, depends only of the circle radius.
It is independent of the characteristics of the diffusion walk~:
the mean free path $\lambda$ (average distance between two
scattering events, i.e. between two direction-changes),
and the single-scattering phase function
$p({\bf u_s},{\bf u_i})$ (probability density function of the
scattering direction ${\bf u_s}$ for an incident direction ${\bf u_i}$).
Furthermore, this average time scales as $R/v$,
which means that the average trajectory-length $<L>$ scales as $R$.
The average trajectory-length would therefore be the same for 
different experiments with any insect species - or for any type of
diffusive corpuscular motion.

There are two reasons why this observation may initially 
sound counter-intuitive.
The first reason is that the average length of diffusion trajectories
between two points is known to scale approximately
as $d^2/\lambda$ where $d$ is the
distance between the two points \footnote{ This scaling would  be
exact for an infinite domain in the limit $d >>\lambda$ \cite{Morse53}}.
For shorter mean free paths,
because of more frequent direction-changes,
it should be required that ants cover more distance to go from one
point to another (see Fig. 1). One could therefore expect an
increase of $<L>$ when reducing the mean free path.

The second reason is the following.
When the mean free path is very small compared to $R$,
most ants have an exit point very close to their entry point.
This means that for analysis of short trajectories, the circle
curvature may be forgotten.
Consequently, there is an homothety of trajectories
corresponding to a mean free path $\lambda$ and say a mean free path
$\lambda/2$ (see Fig. 3). This should imply that $<L>$ decreases
when reducing $\lambda$, which is the opposite conclusion of that
corresponding to the first way of reasoning.

Obviously these two images are both meaningful, but the corresponding
processes are simultaneously under way and compensate each other
exactly. The first image concerns long trajectories (large numbers of
scattering events before exit) and the second one concerns short
trajectories (exit after a few scattering events).
Why this exact compensation occurs is not obvious, but a simple
proof can be proposed on the basis of transport theory.

\section{Property}

Let us first generalize the underlying property.
Consider any finite volume $\Omega$ of envelope $\Sigma$ submitted to 
a uniform isotropic particle-incidence. 
$\Omega$ is filled with any non-homogeneous anisotropic 
perfectly diffusing medium.
For a statistical event in which a particle enters $\Omega$ through 
$\Sigma$, the total trajectory length $L$ is defined as the
length of the multiple scattering trajectory from entry point to first
exit through $\Sigma$. Then the average value of $L$ is independent
of both the mean free path and single-scattering phase function
fields\footnote{For two-dimension geometries Eq. 1 becomes
$<L>=\pi\frac{S}{P}$ where $S$ is the surface of the considered system
and $P$ its perimeter.}~:
\begin{equation}
<L>=4\frac{\Omega}{\Sigma}
\end{equation}
Note that this relationship may be seen as a generalization of
the well known mean-chord-length property \cite{Case67} that
corresponds to Eq. 1 in the limit of an infinite mean free path
(straight-line trajectories) \footnote{ The assumption of isotropic incidence
is formally equivalent to a cosine angular distribution. For instance, if
$\Sigma$ is an infinite planear slab of thickness $e$, for straight-line trajectories
$<L>$ writes $\int_{0}^{\pi/2} 2 cos\theta \; sin \theta \; l( \theta ) \; d \theta$ 
with $l(\theta)=\frac{e}{cos \theta}$, leading to $<L>=2 e$, which is 
compatible with the mean-chord-length property with 
$\frac{\Omega}{\Sigma}= \frac{e}{2}$.
}.

{\bf Proof~:}
The property addressed is purely geometric. In particular, if two sets
of particles are considered, with same mean free path and same single-scattering
phase function, $<L>$ will be identical, independently of the particle speed.
This means that if it can be shown that the property is true for constant-speed
diffusion, then the property is true for any particle-diffusion process.
Consequently, in order to establish the validity of Eq. 1, it is rigorously
sufficient to derive a proof in the restricted case
of a constant-speed displacement.

On this basis, the starting point is the following~:
one way of creating a uniform isotropic particle incidence
is to consider that $\Omega$ is part of a larger diffusing system in a state
of statistical equilibrium. The specific intensity $f({\bf x},{\bf u})$
(number of particles passing at a location ${\bf x}$, in a direction
${\bf u}$, per unit time, per unit normal area and per unit solid angle)
is therefore uniform and isotropic~: $f({\bf x},{\bf u})=f_0$.
From this equilibrium state, let us imagine
that a uniform particle-absorption, of absorption coefficient $\mu_a$ (inverse of the absorption 
mean-free-path), is added
to the physics of $\Omega$.
This means that the number of absorbed particles at a each location ${\bf x}$,
per unit time and per unit volume, is
$\mu_a v \int_{4\pi} f({\bf x},{\bf u}) d{\bf u}$ where $v$ is the particle speed.
This also means that for any particle traveling a distance $L$ through $\Omega$, the
probability that an absorption occurs before exit is $1 - \exp(-\mu_a L)$. 
In the stationary state,
two ways of expressing the total absorption rate inside $\Omega$ are therefore
available~: the first is to integrate the local absorption rate over the volume;
the second is to consider the incident particle flux at the boundary
$\pi v f_0 \Sigma$ (the factor $\pi$ appears because of isotropy) and to
integrate the absorptions over all possible trajectories trough $\Omega$.
This leads to the following equation~:
\begin{equation}
\int_{\Omega}
\left[
\mu_a v \int_{4\pi} f({\bf x},{\bf u}) d{\bf u}
\right]
d{\bf x}
=
\pi  v f_0 \Sigma \int_0^{\infty} \left[ 1 - \exp(-\mu_a L) \right] p(L) dL
\end{equation}
where $p(L)$ is the probability density function of the trajectory length.

Let us now divide by $\mu_a$ both sides of Eq. 2 and take the limit
$\mu_a \rightarrow 0$. At this limit, the system is again purely diffusing
and statistical equilibrium is satisfied~:
$f({\bf x},{\bf u})=f_0$. The left hand side of the equation therefore becomes
proportional to $\Omega$. On the right hand side, the limit
$\frac{1}{\mu_a}[1 - \exp(-\mu_a L)] \rightarrow L$
turns the integral into the average trajectory length
$<L>=\int_0^{\infty} L \ p(L) dL$. Altogether we get~:
\begin{equation}
4 \pi v f_0 \Omega = \pi  v f_0 \Sigma <L>
\end{equation}
which leads to Eq. 1.

\section{Interest}

Such an invariance property ($<L>$ remaining constant when changing the random-walk
characteristics) offers a challenging opportunity to test available
physical pictures of particle diffusion. As illustrated in the introduction,
only an exact compensation of physical trends associated with long and short trajectories
can explain Eq. 1, but this compensation is hard to appreciate intuitively.

Outside the intrinsic theoretical interest of such a compensation effect,
the property presented might allow
a renewed look at the open question of modeling the statistics of short-length
diffusion trajectories. The statistics of long trajectories has been modeled
successfully thanks to the assumption that the number of scattering events along each
such trajectory is high, which allows the use of averaging relations\cite{Morse53}.
As far as short trajectories are concerned, less successful work is
reported, mainly because the corresponding statistical problem is of a much
higher complexity\cite{Freund88,Yoo90,Ray92}.
Of course, Eq. 1 as such solves none of the corresponding problems,
but the fact that constraint relationships can be exhibited between the statistics of
short trajectories and long trajectories might be of direct interest for researchers
in these fields.

As far as practical applications are concerned, despite the very wide
range of physics and biology problems in which analysis of particle-diffusion
trajectories are required, the following objection may be formulated~: application
of Eq. 1 requires that the considered system be purely diffusing and that the
particle incidence be uniform and isotropic, both conditions that are sufficient to
ensure that the particle distribution tends to statistical equilibrium, which turns
the problem into one of limited interest. Present-day questions that involve
detailed analysis of diffusion processes concern non-equilibrium conditions
and Eq. 1 may therefore appear of little practical use.

One way of answering this objection is to refer to all configurations in which
symmetry considerations allow to Eq. 1 to be applied under non-equilibrium conditions.
Let us consider again the biology example used in the introduction. All entry points
on the circle play a similar role, which in turn allows us to consider experiments
in which only one entry location is offered. Such experiments are typical of 
conditions
where the proposed property is applicable, although the distribution of ants inside
the circle does not fulfill conditions for statistical equilibrium.
Another more practical example
along the same line may be seen in the field of radiative transfer for planetary
atmospheres. The basic assumption is commonly that atmospheres may be divided into
homogeneous plane-parallel horizontal layers. There are numerous conditions in which
the assumption may be made that incident photons on each side are distributed
quasi-isotropically. This is typically the case when photon sources are distributed within
the atmosphere (as for infrared radiation in the earth's atmosphere, where collimated
solar sources can be neglected \cite{GoodyYung89} ). 
The upward and downward photon fluxes being distinct,
each layer is not in a state of statistical equilibrium. However, for symmetry reasons,
Eq. 1 may be applied separately to upward and downward photon incidences.
Very similar illustrations could be found in neutron transport as well as in
atomic charged-particle transport.

Another class of problems where Eq. 1 may be useful for analysis of non-equilibrium
conditions is the following~: problems where uniformity and isotropy
of particle incidence is effectively satisfied (strictly or via symmetry assumptions),
but where the system is not purely diffusing. For example, considering photon migration
in turbid media, mainly for medical-imaging applications, it was shown that the photon
propagation process could be separated into an absorption dependent part and a pure
diffusion part\cite{Perelman94,Perelman95}. We used such a physical picture in the above presented
proof, when writing the total absorption rate inside the system (in the case of a uniform
absorption coefficient $\mu_a$) as
$A=\pi  v f_0 \Sigma \int_0^{+\infty} [1-exp(-\mu_a L)] \ p(L) dL$.
In this expression, $L$ is indeed the trajectory length of photons across the system
as if no absorption occurred and we can state that $<L>=4\frac{\Omega}{\Sigma}$.
In the limit of a weak absorption for instance (optically thin absorption),
the total absorption rate may therefore be approximated (linearizing the exponential) as 
\begin{equation}
A \approx \pi  v f_0 \Sigma \int_0^{+\infty} \mu_a L \ p(L) dL
= \pi  v f_0 \Sigma \mu_a <L> = 4 \pi \Omega v f_0 \mu_a
\end{equation}
This property allows radiative transfer exchanges to be easily quantified inside
systems of very complex geometries, as is already commonly performed in
thermal-engineering in the limit of non-scattering radiation (thanks to the
mean-chord-length property)\cite{SiegelHowell92}. Equation 4 establishes that such
engineering techniques may be rigorously extended to configurations with non-negligible photon
scattering effects.

Finally, Eq. 1 may be experimentally used to test the relevance of diffusion
assumptions, in particular in biology contexts where suspicions of external
orientation or internal marking effects regularly arise.
This is possible as  soon as experimental techniques are implemented, allowing
individual paths to be tracked and stored automatically, which is undeniably
a common trend in present-day research into animal behavior.
%
%
%
%._._._._._._._._._._._._._._._._._._._._._._._._._._._._._._.
%

\newpage

\begin{figure}
%\onefigure{cercle.eps}
\begin{center}
\includegraphics[width=0.8\textwidth,, angle=0]{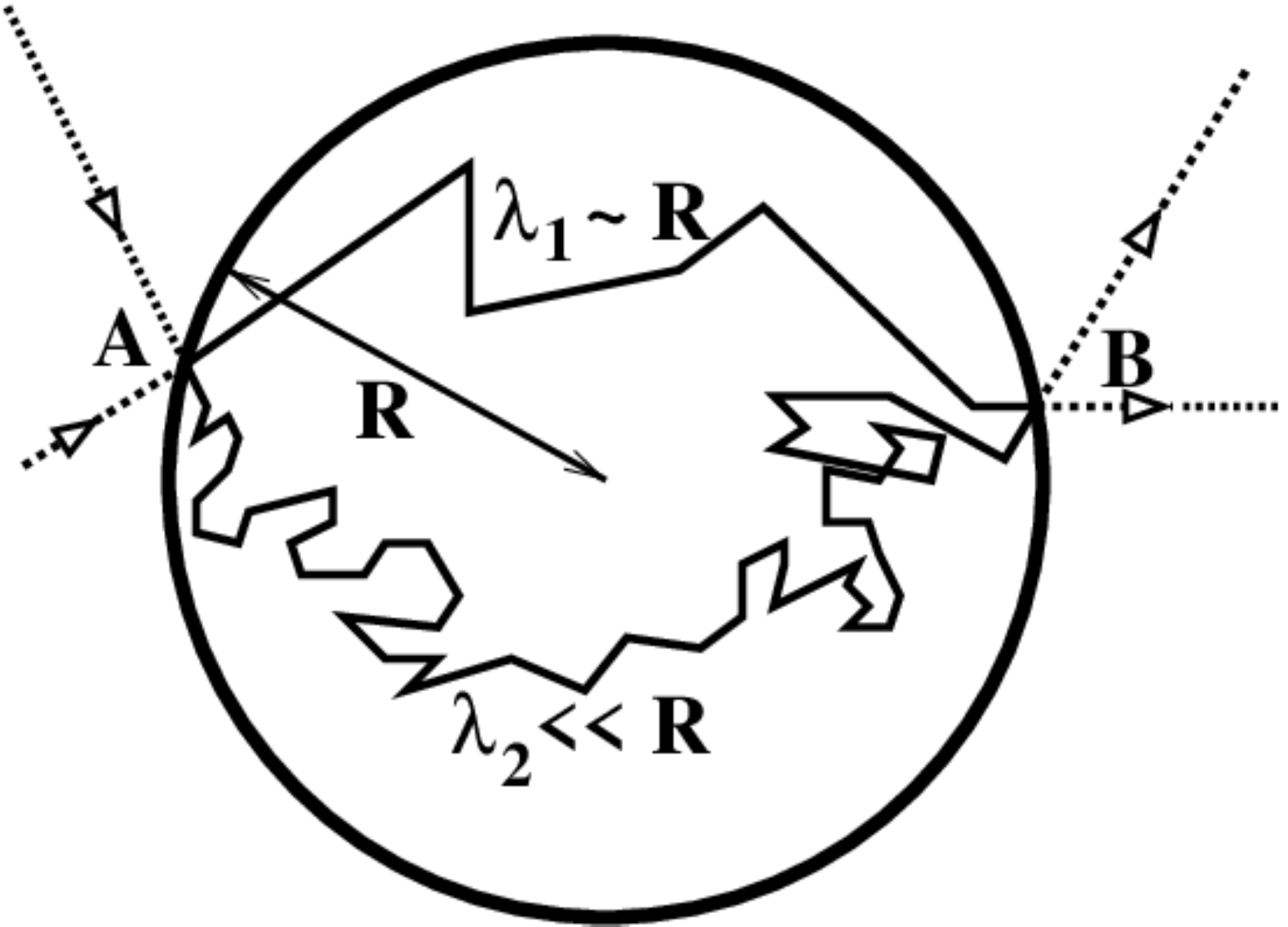}
\end{center}
\caption{ {\bf Random walk illustration -}
Trajectories of diffusing particles inside a circle of radius $R$, from
entry point $A$ to exit point $B$. The first trajectory is typical of a
particle with a mean free path $\lambda_1$ of the order of $R$, whereas
the second trajectory corresponds to a mean free path $\lambda_2 < \lambda_1$;
this illustrates the fact that the trajectory lengths inside the circle, $L_1$
and $L_2$ respectively, should averagely verify $L_1 < L_2$.}
\label{f.1}
\end{figure}

\newpage
\begin{figure}
%\onefigure{Lbar.ps}
\begin{center}
 \includegraphics[width=0.6\textwidth, angle=-90]{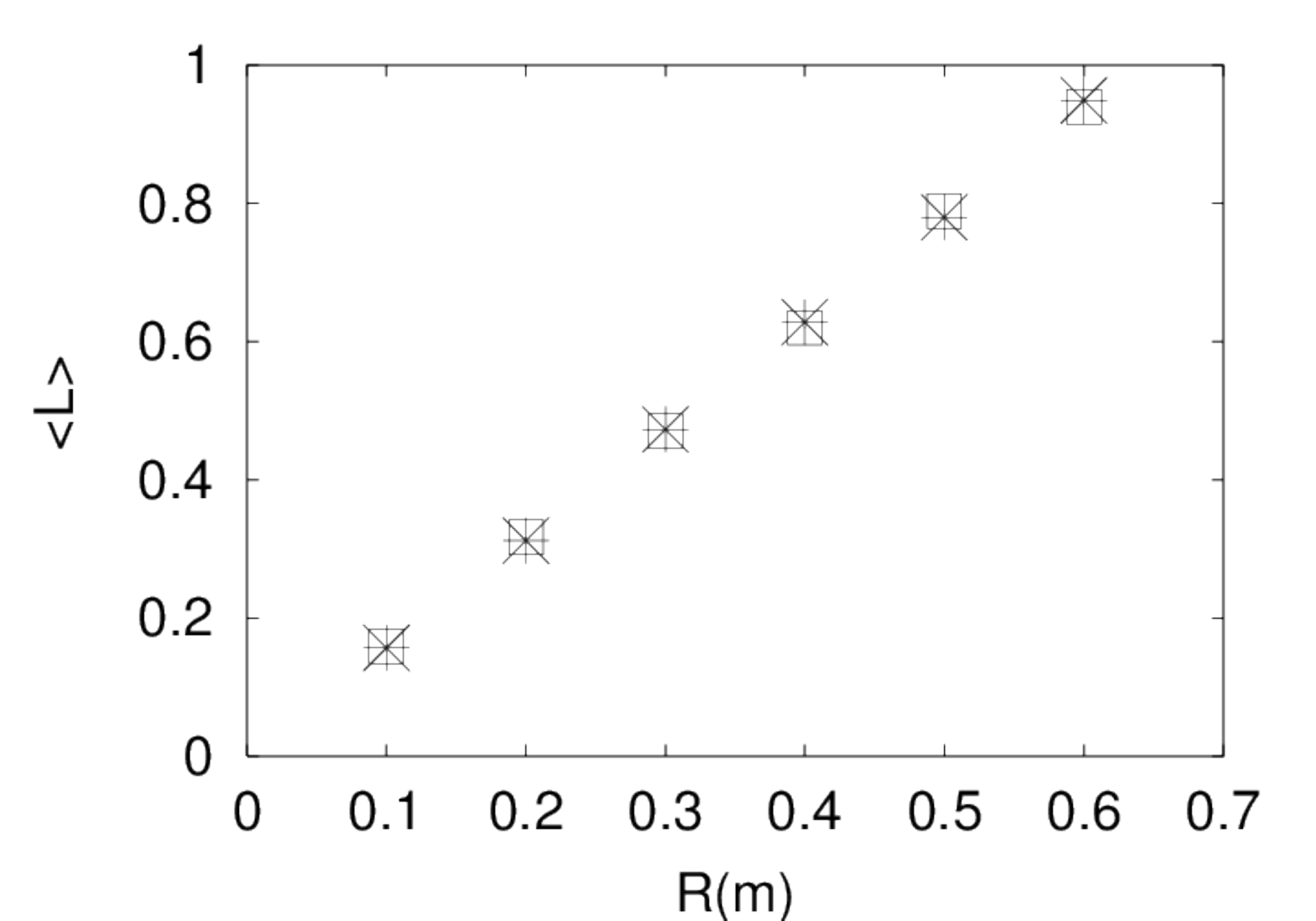}
\end{center}
\caption{{\bf Monte Carlo simulations -}
Simulated average trajectory-length inside a circle of radius $R$ for two 
insects with distinct random-walk characteristics~: $\ast$  mean-free-path
$\lambda=5. \ 10^{-3} m$ (measured for the ant species {\it Messor sancta} L.);
$\square$ mean-free-path $\lambda=5. \ 10^{-2} m$
(roughly estimated for the cockroach species {\it Blattella germanica} L.).
In both simulations,
the asymmetry parameter of the single scattering phase function is $g=0.5$. }
\label{f.2}
\end{figure}

\newpage

\begin{figure}
%\onefigure{reflexion.eps}
\begin{center}
\includegraphics[width=0.8\textwidth, angle=0]{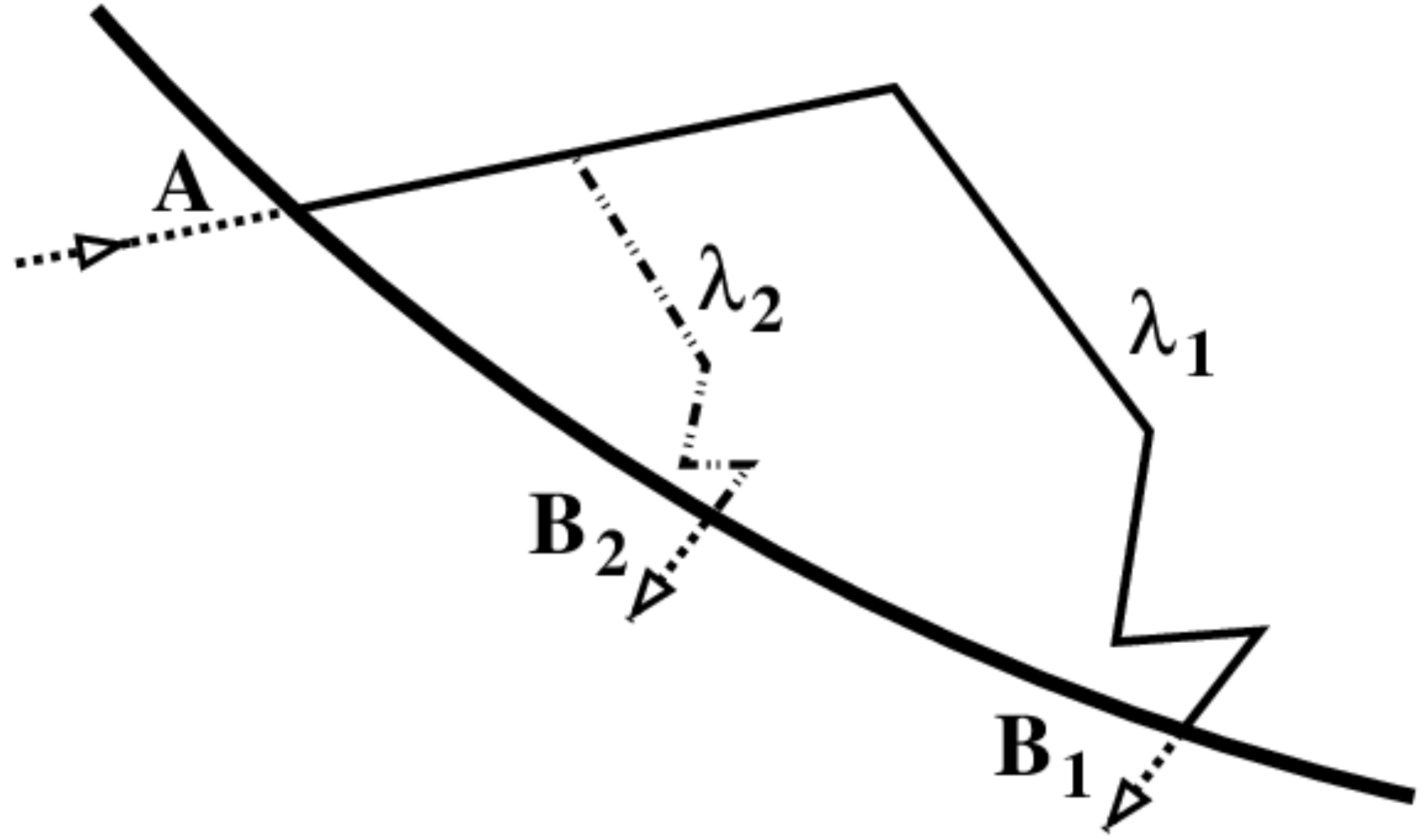}  
\end{center}

\caption{ {\bf Trajectory homothety -}
Illustration of the quasi-homothety of short trajectories.
The second trajectory corresponds to a mean free path half as long as
that of the first trajectory ($\lambda_2 = \lambda_1/2$), which for same
statistical choices, and neglecting circle curvature,
leads to a trajectory length half as long ($L_2 \approx L_1/2$). 
     }
\label{f.3}
\end{figure}

%Paper text.
%See fig.~\ref{f.1}, table~\ref{t.1} and eq.~(\ref{e.1}).
%See also~\cite{b.a,b.b}.
%\begin{equation}
%\label{e.1}
%0\neq1
%\end{equation}

%\begin{figure}
%\onefigure{epl-template.eps}
%\caption{Figure caption.}
%\label{f.1}
%0\end{figure}

%\begin{table}
%\caption{Table caption.}
%\label{t.1}
%\begin{center}
%\begin{tabular}{lcr}
%first  & table & row\\
%second & table & row
%\end{tabular}
%\end{center}
%\end{table}

%\acknowledgments
%Paper text.

\end{document}